\documentclass[twocolumn,showpacs,preprintnumbers,amsmath,amssymb]{revtex4}

\usepackage{graphicx}
\usepackage{dcolumn}
\usepackage{bm}

\begin{document}


\title{Fourier Analysis of Ghost Imaging} 

\author{Honglin Liu}\thanks{E-mail: hlliu4@gmail.com.}%

\author{Jing Cheng}
\author{Yanfeng Bai}
\author{Shensheng Han}%
\affiliation{Key Laboratory for Quantum Optics and the Center for
Cold Atom Physics of CAS, Shanghai Institute of
Optics and Fine Mechanics.\\
Chinese Academy of Sciences, P.O.Box 800-211, Shanghai, 201800,
China
}%

\date{\today}

\begin{abstract}
Fourier analysis of ghost imaging (FAGI) is proposed in this paper
to analyze the properties of ghost imaging with thermal light
sources. This new theory is compatible with the general
correlation theory of intensity fluctuation and could explain some
amazed phenomena. Furthermore we design a series of experiments to
testify the new theory and investigate the inherent properties of
ghost imaging.
\end{abstract}

\pacs{42.50.Dv,03.65.Ud,42.60.st}
 \maketitle

Conventionally, coherent, at least partially coherent,
illumination is required to image the phase detail of an object
[1], and the theoretical limit of the resolution is determined by
the wavelength $\lambda$  of the light beam. Since the coherent
sources in the ultra-violation and even shorter frequency regions
are still not existed, we could not obtain a higher resolution by
using a coherent light with a shorter wavelength. In 1994,
Belinsky and Klyshko [2] found that, by exploiting the spatial
correlation between two entangled photons generated by parametric
down conversion (PDC), ghost imaging could be realized with
entangled incoherent light. Due to the possibility to improve the
resolution, ghost imaging is bestowed with great potential in
quantum metrology, quantum holography and quantum
lithography,\emph{etc}.

The nature of ghost imaging leads to many interesting debates
about the necessity of the entanglement [3-14]. Presently, the
idea that both classical thermal sources and quantum entangled
beams can be used for ghost imaging and ghost diffraction is
widely accepted.

In order to realize practical use of the ghost imaging, further
investigation about the properties of ghost imaging is stringent.
 Gatti \emph{et al.} compared the visibility in the classical
and quantum regimes, and pointed that the visibility of classical
regime increases as the mean photon number of per mode in thermal
state increases [9]. D'Angelo \emph{et al.} compared the
resolution of quantum and classical ghost imaging [15]. Meanwhile,
Ferri \emph{et al.} realized the high-solution of ghost imaging
and ghost diffraction with thermal light [16], and they even tried
ghost imaging with homodyne detection [17]. Bache \emph{et al.}
analyzed the spatial average technique which would make the
imaging bandwidth of the reconstructed diffraction pattern
virtually infinite and the correlation convergence rate faster
[18]. Cheng \emph{et al.} have presented theoretical analysis on
the noise of ghost imaging with entangled photons [19] and
pseudo-thermal light fields [20]. Most recently our group
demonstrated lensless Fourier-transform ghost imaging with both
amplitude only and phase only object by using classical incoherent
light [21].

There are still some questions unresolved, such as what limits the
complexity of the object, why the visibility decreases as the
transmission increases. In this paper we propose a new theory,
Fourier analysis of ghost imaging (FAGI), to answer these
questions.

For an ideal infinite and uniform thermal source located at $z=0$,
the field at a point in the source is presented by
$U_{0}(x_{0},y_{0},t)=\int\int_{\pm\infty}dk_{x}dk_{y}A\exp
[i(k_{x}x_{0}+k_{y}y_{0}-\omega t)]$ . When
$k_{x}^{2}+k_{y}^{2}\geq(\frac{\omega}{2\pi})^{2}$,$k_{z}$ is
imaginary, and the amplitude of this evanescent component decays
exponentially as the distance $z$ increases, so the integral
region can be confined in $k_{i}\in(-k,k),i\in(x,y)$. Since the
sources used in experiments are always finite, which means $k_{x}$
and $k_{y}$ are not uniformly distributed, the typical Gaussian
distribution is adopted. For simplicity $k_{x}$ is assumed to be
independent of $k_{y}$, and the amplitude $A$ is a time variant
and random distributed, so the field is written as
$U_{0}(x_{0},y_{0},t)=\int\int_{\pm\infty}dk_{x}dk_{y}\exp(-\frac{k_{x}^{2}+k_{y}^{2}}{\sigma^{2}})A_{\vec{k}}(t)\exp
[i(k_{x}x_{0}+k_{y}y_{0})]$.

Now we analyze the paradigmatic example of the imaging system
given in Ref.[22], except that the source is replaced by a thermal
source. Here the setup of the object arm is an f-f system with the
object close to the beam splitter (BS). For simplicity the
output-plane of the BS is selected as the source, the fields on
these two planes are shown as

\begin{eqnarray}\label{1}
U(x_{0},y_{0},t)&=&\frac{1}{2}\int\int_{\pm\infty}dk_{x}dk_{y}\exp(-\frac{k_{x}^{2}+k_{y}^{2}}{\sigma^{2}})\nonumber\\&&A_{\vec{k}}(t)\exp
[i(k_{x}x_{0}+k_{y}y_{0})]
\end{eqnarray}
\begin{eqnarray}\label{2}
V(x_{0}^{'},y_{0}^{'},t)&=&\frac{1}{2}\int\int_{\pm\infty}dk_{x}dk_{y}\exp(-\frac{k_{x}^{2}+k_{y}^{2}}{\sigma^{2}})\nonumber\\&&A_{\vec{k}}(t)\exp
[i(k_{x}x_{0}^{'}+k_{y}y_{0}^{'})]
\end{eqnarray}
and the object transmitting function is $t(x_{0},y_{0})$.
Neglecting the finite pupil of the lens, through the Fourier
Transform of the lens, the spatial frequency distribution at the
focal plane of the object arm is
\begin{eqnarray}\label{3}
U(x_{f},y_{f},t)&=&\frac{1}{2}\int\int_{\pm\infty}dk_{x}dk_{y}\exp(-\frac{k_{x}^{2}+k_{y}^{2}}{\sigma^{2}})\nonumber\\&&A_{\vec{k}}(t-\frac{2f}{c})T[(\frac{2\pi}{\lambda
f}x_{f}-k_{x}),(\frac{2\pi}{\lambda f}y_{f}-k_{y})]\nonumber\\
\end{eqnarray}

Here, the constant phase, which has no sensible influence on the
result,is neglected. Because of the incoherence of the thermal
source, a fixed pixel detector collects a certain range of spatial
frequency components. Comparing with a single frequency component
in the coherent illumination case, it is easy to understand why
the ghost imaging can be realized with only a pixel detector in
the object arm.

In this system, the reference arm is a 2f-2f setup, and the
transmitting function
$h(x_{2f},y_{2f},x_{0}^{'},y_{0}^{'})=\delta(x_{2f}-x_{0}^{'},y_{2f}-y_{0}^{'})$,
so the field distribution on the detector plane is
\begin{eqnarray}\label{4}
V(x_{2f},y_{2f},t)&=&\frac{1}{2}\int\int_{\pm\infty}dk_{x}dk_{y}\exp(-\frac{k_{x}^{2}+k_{y}^{2}}{\sigma^{2}})\nonumber\\&&A_{\vec{k}}(t-\frac{4f}{c})\exp [i(k_{x}x_{2f}+k_{y}y_{2f})]\nonumber\\
\end{eqnarray}

The intensity fluctuation correlation function is
$G(x_{1},y_{1},x_{2},y_{2})=|\Gamma(x_{1},y_{1},x_{2},y_{2})|^{2}$,
and the second order correlation function
$\Gamma(x_{f},y_{f},x_{2f},y_{2f})=\left<U(x_{f},y_{f},t)V^{\ast}(x_{2f},y_{2f},t)\right>$.
On the assumption of the independence of the amplitude of
different wave vectors, $\langle
A_{\vec{k}}(t-\frac{2f}{c})A_{\vec{k}^{'}}(t-\frac{4f}{c})\rangle=0$,
when $\vec{k}\neq\vec{k}^{'}$, the second order correlation
function is simplified as
\begin{widetext}
\begin{eqnarray}
\Gamma(x_{f},y_{f},x_{2f},y_{2f})=
\frac{1}{4}\int\int_{\pm\infty}dk_{x}dk_{y}\langle
A_{\vec{k}}(t-\frac{2f}{c})A_{\vec{k}}(t-\frac{4f}{c})\rangle\exp(-2\frac{k_{x}^{2}+k_{y}^{2}}{\sigma^{2}})\nonumber\\T[(\frac{2\pi}{\lambda
f}x_{f}-k_{x}),(\frac{2\pi}{\lambda
f}y_{f}-k_{y})]\exp[-i(k_{x}x_{2f}+k_{y}y_{2f})] \label{eq:wideeq}
\end{eqnarray}
\end{widetext}

Within the coherent time,  $\langle
A_{\vec{k}}(t-\frac{2f}{c})A_{\vec{k}^{'}}(t-\frac{4f}{c})\rangle$
is a const. When $x_{f}=0$ ,$y_{f}=0$ ,eq.(5) is presented as:
\begin{widetext}
\begin{eqnarray}\label{6}
\Gamma(0,0,x_{2f},y_{2f})=\frac{1}{4}I\int\int_{\pm\infty}dk_{x}dk_{y}\exp(-2\frac{k_{x}^{2}+k_{y}^{2}}{\sigma^{2}})T(
-k_{x},-k_{y})\exp[-i(k_{x}x_{2f}+k_{y}y_{2f})]
\end{eqnarray}
\end{widetext}
this equation is a Fourier Transform, where the object spectrum
multiplies a Gaussian function, which causes a increasing loss for
high frequencies. According to the properties of Fourier
Transform, a convolution of the object image and a Gaussian
function is obtained:
\begin{widetext}
\begin{eqnarray}\label{7}
\Gamma(0,0,x_{2f},y_{2f})=\frac{\sigma^{2}I}{16}t(-x_{2f},-y_{2f})\otimes\exp(-\sigma^{2}\frac{x_{2f}^{2}+y_{2f}^{2}}{8})
\end{eqnarray}
\end{widetext}
The presentation of the Gaussian spreading function contributes a
background to the object image, which explains the occurrence of
the inherent noise in the second order intensity fluctuation
correlation. From this equation it easy to deduce that when the
transmission area increases, the background increases while the
visibility decreases.

When $x_{f}\neq0$ ,$y_{f}\neq0$ , the second order correlation
function is
\begin{equation}
\Gamma(x_{f},y_{f},x_{2f},y_{2f})\propto\exp[-i\frac{2\pi}{\lambda
f}(x_{f}x_{2f}+y_{f}y_{2f})]\Gamma(0,0,x_{2f},y_{2f})
\end{equation}

If the pixel detector is replaced by a bucket
detector in the object arm, the corresponding result will be
\begin{widetext}
\begin{eqnarray}\label{9}
\int\int
dx_{f}dy_{f}\Gamma(x_{f},x_{f},x_{2f},y_{2f})\propto\int\int
dx_{f}dy_{f}\exp[-i\frac{2\pi}{\lambda
f}(x_{f}x_{2f}+y_{f}y_{2f})]\Gamma(0,0,x_{2f},y_{2f})
\end{eqnarray}
\end{widetext}

Similar results can be obtained with only coefficient difference.

\begin{figure}
\includegraphics[height=8cm]{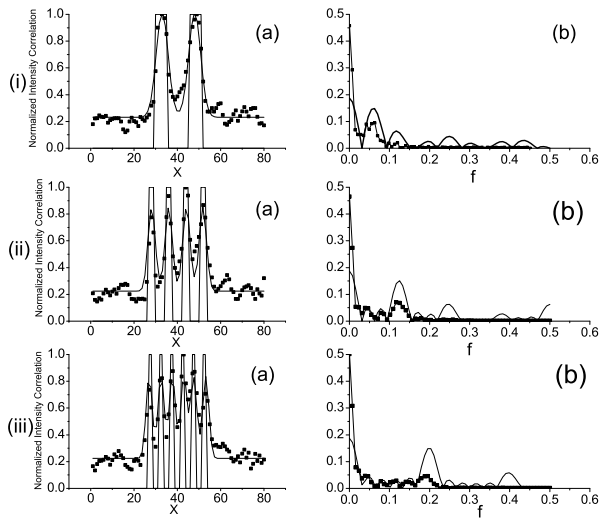}
\caption{\label{fig:epsart} (a)the image of different objects and
(b)the corresponding spatial frequency spectrum. (i)two slits, the
width is 300 $\mu m$, and the distance is 900 $\mu m$. (ii)four
slits, width 150 $\mu m$, and distance 450 $\mu m$, (iii)six
slits, width 100 $\mu m$, and distance 300 $\mu m$.}
\end{figure}

We have designed several experiments to demonstrate these
deductions. Since the source we used is a pseudo thermal source,
the best resolution is limited by the coherence length $l_{c}$ ,
simply by replacing $\lambda$  with $l_{c}$, all results obtained
in previous discussions are still correct. In the experiment, the
coherence length of the source is 75$\mu m$ .

The first experiment is to investigate the frequency response of
ghost imaging. The objects have the same duty cycle but different
periods, and we get the image and the corresponding spatial
frequency distribution of each object.

In Fig.1, the black bold line in each (a) is the object, the
disperse squares are the normalized intensity fluctuation
correlation, and the fitting curve is the image. Each (b) is the
corresponding spatial frequency distribution of (a).

In this experiment we choose the ratio of the slit width and the
distance $a:d=1:3$ to make sure the image reconstructed by low
frequency has no direct current background. Moreover, comparing
the first-order component of the image to the zero-order component
of the object in Fig.1 (i), the ratio is bigger than 1/2, which
implies no direct current background of zero-order component, so
the background noise is the contribution of the convolution of
Gaussian spread function. As the frequency increases, the ratio of
image first-order component to object first-order component
decreases, so the effective information decreases. Results are
shown in Table.1 and Fig.2.

\begin{figure}
\includegraphics[height=7cm]{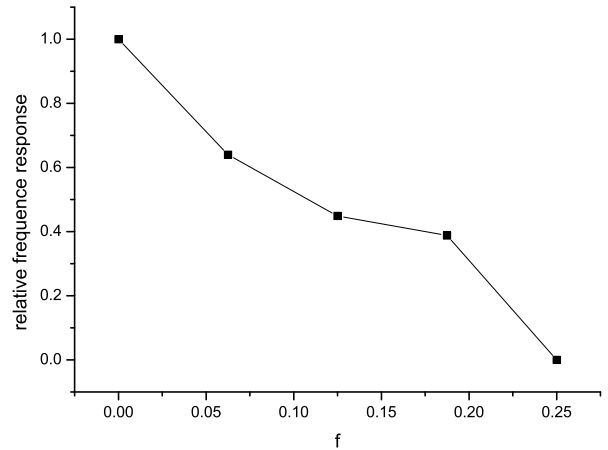}
\caption{\label{fig:epsart} The frequency response curve of
Table.I.}
\end{figure}

\begin{table}
\caption{\label{tab:table1}the relative frequency response rate in
ghost imaging. (the coherence length is 75 $\mu m$) }
\begin{ruledtabular}
\begin{tabular}{lcr}
CSL\footnote{Character Spatial Length.}&FFC\footnote{Frequency of First-order Component.}&RFR\footnote{Relative Frequency Response.}\\
\hline
infinity & 0 & 1\\
300 & 0.0625 & 0.64\\
150 & 0.125 & 0.449\\
100 & 0.1875 & 0.388\\
75 & 0.25 & 0\\
\end{tabular}
\end{ruledtabular}
\end{table}

In Fig.2 the frequency response curve has an approximate Gaussian
form, as given in the theoretical result Eq.(6).

\begin{figure}
\includegraphics[height=11cm]{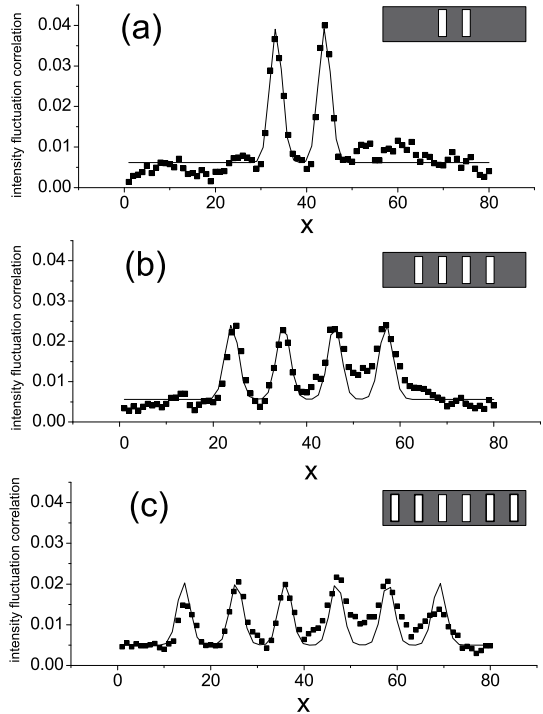}
\caption{\label{fig:epsart} The experiment result of objects with
different transmission area. (a) the image of two-slit, (b)
four-slit and (c) six-slit.}
\end{figure}

In the following experiment we demonstrate the relationship
between the visibility and the transmission area. Here the objects
have the same period parameters but different number of period,
which suggests the spatial spectrums of each object have similar
distribution and response. In eq.(7), the transmission area is
composed by many points, this area can be expressed by a comb
function, each point convoluted with a Gaussian spread function,
and the image is the summation of all Gaussian spreading
functions. As the transmission area increases, more points
contribute to the background that directly makes the visibility
decrease. The experiment result is shown in Fig.3.

The period $d$ is 600$\mu m$, and the width $a$ of the slit is
200$\mu m$. From this experiment it seems reasonable to say the
visibility decreases with the increase of the transmission area.

Based on the second experiment we suppose that when the
transmission area is large enough, the image is submerged in the
inherent noise. Here we design a two-slit and its reverse, the
objects are shown in Fig.4.

\begin{figure}
\includegraphics[height=7cm]{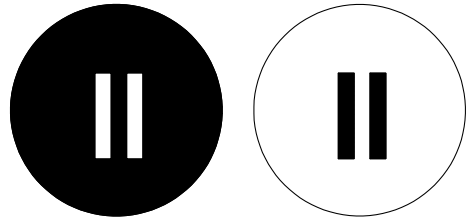}
\caption{\label{fig:epsart} (a)the object and (b)its reverse.}
\end{figure}

in experiment, the object (a) is easy to be imaged, but under the
same condition, its reverse (b) can not be imaged.

From the above experiment results, we conclude that the Fourier
analysis theory presented in this paper can interpret the
principle of ghost imaging in different systems, and resolve some
difficulties of intensity correlation function.

In summary, we propose a new theory to clarify the principle of
ghost imaging, which is compatible with the generally used
intensity fluctuation correlation theory, and can be used to
explain the phenomena we realized in experiments. Moreover, the
analysis of ghost imaging with a classical thermal source can be
generalized to the case of an entangled source, which remains for
the future work.

\begin{acknowledgments}
Support from the National Natural Science Foundation of China
(60477007 and 10404031), Shanghai Rising-Star Program and the
Shanghai Optical-Tech Special Project (034119815) is acknowledged.
\end{acknowledgments}

\end{document}